\documentclass[11pt]{article}
\usepackage[style=apa, backend=biber, url=true, doi=true]{biblatex}

\usepackage{amsmath}
\usepackage{amssymb}
\usepackage{amsthm}
\usepackage{bbm}
\usepackage{bm}

\usepackage{graphicx}

\usepackage{apalike}
\usepackage{booktabs}
\usepackage{caption}
\usepackage{subcaption}
\usepackage{a4wide}

\usepackage{multirow}
\usepackage{siunitx}

\usepackage{pdflscape}
\usepackage{afterpage}
\usepackage{capt-of}
\usepackage{hyperref}
\usepackage{lipsum}

\usepackage{changes}
\usepackage{xcolor}
\newcommand{\stkout}[1]{\ifmmode\text{\sout{\ensuremath{#1}}}\else\sout{#1}\fi}
\setdeletedmarkup{\stkout{#1}}

\theoremstyle{definition}

\providecommand{\keywords}[1]{\textbf{\textit{Keywords -- }} #1}
\providecommand{\jel}[1]{\textbf{\textit{JEL -- }} #1}

\renewbibmacro{in:}{}
\DeclareFieldFormat[article,periodical]{title}{“#1”} 
\DeclareFieldFormat{journaltitle}{\mkbibemph{#1}}    

\begin{document}

\title{Price predictability in limit order book with deep learning model}

\author{Kyungsub Lee\footnote{Corresponding author, \href{mailto:ksublee@yu.ac.kr}{ksublee@yu.ac.kr}, Department of Statistics, Yeungnam University, Gyeongsan, Gyeongbuk 38541, Korea}\\ }

\maketitle

\begin{abstract}
This study explores the prediction of high-frequency price changes using deep learning models. 
Although state-of-the-art methods perform well, their complexity impedes the understanding of successful predictions. 
We found that an inadequately defined target price process may render predictions meaningless by incorporating past information.
The commonly used three-class problem in asset price prediction can generally be divided into volatility and directional prediction. 
When relying solely on the price process, directional prediction performance is not substantial. 
However, volume imbalance improves directional prediction performance.
\end{abstract}


\jel{G17, C88}

\section{Introduction}

Recent research has focused on high-frequency price change prediction using advanced deep learning techniques and rich information from limit order book (LOB) data. 
Despite the success of these deep learning models,
the reasons behind their highly accurate predictions are not well understood.
This lack of explanation raises questions regarding the interpretability of these models.

Our goal is not to propose a new deep learning model or compare existing ones for marginal improvements. 
Instead, we focus on discussing the key factors for meaningful performance in high-frequency price prediction using deep learning.

To understand the prediction process, we categorize predictability into two dimensions, volatility and directional predictability, constituting one of our key observations. 
While the literature often classifies future price changes as up, down, or stable, we further analyze this three-class problem to provide a more comprehensive understanding of the factors that influence price dynamics.

\section{Prediction with deep learning model}

\subsection{Modified return process}

Similar to prior work such as \textcite{zhang2019deeplob}, \textcite{Wang2023}, and \textcite{yin2024forecasting}, we use neural network approaches to anticipate mid-price movements in LOBs.
Our preprocessing method aligns with the standard procedure for predicting future price process.
In many studies, including ours, standardization processes are applied to bids, ask prices, and volumes to enhance performance.

Let $m$ denote the standardized mid-price process, calculated using the previous five days' data.
Time $t$ is indexed by integers,
representing the moments at which changes in the LOB occur.
Let 
$$ p_k(t) = \frac{1}{k}\sum_{i=0}^{k-1} m(t - i), \quad f_k(t) = \frac{1}{k} \sum_{i=1}^{k} m(t + i),$$
the averages of standardized mid-prices for the past and future, respectively.
The modified return is defined as 
\begin{equation}
r_{k,k'}(t) = \frac{f_{k'}(t) - p_k(t)}{p_k(t)}. \label{Eq:r}
\end{equation} 
If $k = k'$, simply let $r_k = r_{k,k'}$.

Previous studies have used benchmark data, such as FI-2010 \parencite{ntakaris2018benchmark}, which includes LOB data for 10 stocks on the NASDAQ's Nordic exchange in 2010. 
However, this dataset is dated and has lower activity levels than contemporary LOB data. 
Our analysis uses highly active trading data for AAPL (Apple Inc.) stocks in 2022 from Nasdaq TotalView ITCH. 
These data provide detailed order book information, displaying the top 10 bid and ask price levels, share volume at each level, and timestamps for transactions. 
It captures real-time updates on all order additions, modifications, and cancellations.
We use the data only from standard trading hours, 09:30 to 16:00,
which contain approximately 3 million observations\footnote{An example of the dataset is available at \href{https://doi.org/10.6084/m9.figshare.26166238}{DOI 10.6084/m9.figshare.26166238}.}.

\subsection{Neural network model}

For training with LOB data, we use the DeepLOB model by \textcite{zhang2019deeplob} with the structure presented in Table~\ref{tab:model_summary1}.
This model inputs the entire LOB data or a segment, including up to 10 levels of bids, asks, and volumes. 
This yields an $M$-dimensional time series, where $M$ depends on the chosen levels and elements like price and volume. 
It predicts modified returns using the previous 100 observations before time $t$.
The input variables are standardized as mentioned before.

\begin{table}[t]
	\centering	
	\caption{Summary of the DeepLOB model architecture. The configuration of the convolutional layer (Conv2D) and max pooling is as follows: number of filters (only for convolution), filter size represented by $\times$, Zero padding (if it exists), and stride represented by tuple (if it is not (1, 1)). Each convolution layer is followed by LeakyReLU with a negative slope of 0.01}	\label{tab:model_summary1}
	\begin{tabular}{c|c|c}
		\hline
		\multicolumn{3}{c}{Input layer with $100 \times 40$ input dimension per data} \\
		\hline
		\multicolumn{3}{c}{Conv2D(32, $1\times 2$, (1, 2))} \\
		\multicolumn{3}{c}{Conv2D(32, $4\times 1$, Zero)} \\
		\multicolumn{3}{c}{Conv2D(32, $4\times 1$, Zero)} \\
		\multicolumn{3}{c}{Conv2D(32, $1\times 2$, (1, 2))} \\
		\multicolumn{3}{c}{Conv2D(32, $4\times 1$, Zero)} \\
		\multicolumn{3}{c}{Conv2D(32, $4\times 1$, Zero)} \\
		\multicolumn{3}{c}{Conv2D(32, $1\times 10$)}  \\
		\multicolumn{3}{c}{Conv2D(32, $4\times 1$, Zero)} \\
		\multicolumn{3}{c}{Conv2D(32, $4\times 1$, Zero)} \\
		\hline
		\multicolumn{3}{c}{Inception module} \\
		\hline
		Conv2D(64, $1\times 1$, Zero) & 
		Conv2D(64, $1\times 1$, Zero) & 
		Max pooling($3\times 1$, Zero) \\
		Conv2D(64, $3\times 1$, Zero)  &	
		Conv2D(64, $5\times 1$, Zero)  &
		Conv2D(64, $1\times 1$, Zero) \\
		\hline
		\multicolumn{3}{c}{Concatenate} \\
		\hline
		\multicolumn{3}{c}{Dropout layer with rate 0.2} \\
		\multicolumn{3}{c}{LSTM layer with 64 units} \\
		\multicolumn{3}{c}{Dense layer with 3 units and softmax function} \\
		\hline
	\end{tabular}
\end{table}

The prediction problem is structured as a three-class problem, with the target variable $y = r_{k,k'}(t)$ in Eq.~\eqref{Eq:r}.
Instances are classified as 
$$ \text{UP : } y > \alpha, \quad  \text{DOWN : } y < -\alpha, \quad \text{STABLE : } |y| \leq \alpha $$
for some constant $\alpha$
chosen to balance the class distribution..

Following the input layer, the architecture incorporates multiple convolutional layers,
an inception module, dropout, a long short-term memory layer, and an output layer.
Model evaluation occurs on a daily basis throughout 2022.
For each evaluation, the model is trained using data from the preceding 20 days.

\subsection{Target of $r_{20}$ and naive prediction}

First, we employ $r_{20}(t)$ for prediction, 
following previous studies such as \textcite{tsantekidis2017forecasting} and \textcite{zhang2019deeplob}.
Using the entire LOB before $t$ with a time length of 100 as input, 
we evaluate the test accuracy for all trading days in 2022. 
Table~\ref{tab:lob0} provides the annual averages, with standard deviations in parentheses.

\begin{table}[t]
	\centering
	\caption{Daily average and standard deviation of classification metrics based on the entire LOB data setting a target of $r_{20}$ with DeepLOB model and naive prediction}
	\label{tab:lob0}
	\begin{tabular}{c|cc|cc|c}
		\hline
		& \multicolumn{2}{c|}{DeepLOB} & \multicolumn{2}{c|}{Naive prediction} & \\   
		&  Precision & Recall &  Precision & Recall &  Size\\
		\hline
		UP &  0.672(0.071) & 0.680(0.091) & 0.703(0.038) & 0.583(0.076) & 10,233\\
		DOWN &  0.669(0.075) & 0.685(0.091) &  0.703(0.037) & 0.582(0.076) & 10,229\\
		STABLE & 0.595(0.112) & 0.540(0.190) &  0.524(0.123) & 0.696(0.124) & 10,817\\
		\hline
		Accuracy &  \multicolumn{2}{c|}{0.659(0.038)} &  \multicolumn{2}{c|}{0.648(0.036)} & \\
		\hline
	\end{tabular}
\end{table}

The model from Table~\ref{tab:model_summary1}, using the full LOB as input, attains approximately 65.9\% accuracy per Table~\ref{tab:lob0} left panel.
To understand the high level of accuracy, 
we examine the price process, the target of the forecast.
The time series $r_{k,k'}(t)$ with $k > 1$ in Eq.~\eqref{Eq:r} contains overlapping information by definition, making it naturally predictable.

\begin{figure}[t]
	\centering
	\includegraphics[width=0.5\textwidth]{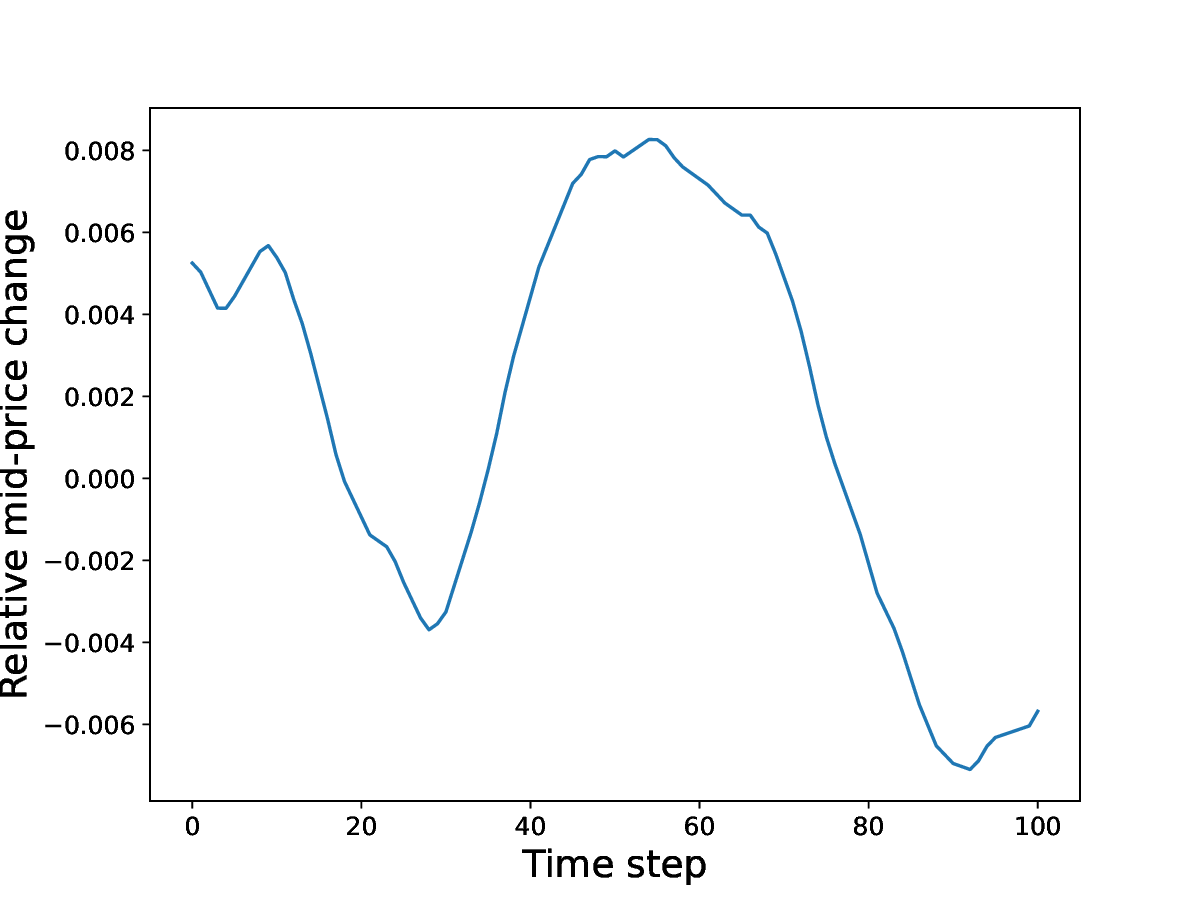}
	\caption{An example of a realized path of $r_{20}$}\label{fig:r_k}
\end{figure}

Figure~\ref{fig:r_k} displays a sample $r_{20}$ path for AAPL on June 29, 2022.  
The $r_{20}$ trajectory in the figure shows a deterministic pattern with less noise, contrasting with usual random price processes.
This suggests that a simple prediction method using the last value to forecast the next could be effective. 
Note that $r_{20}$ inherently contains future information and should not be directly used for naive forecasting.

As $r_{k,k'}(t)$ includes future $k'$ values, 
we, instead, use 
$$ x_{k, k'}(s) = \left\{\begin{array}{ll} 
	r_{k, k'}(s), & \text{for } s \leq t - k' \\
    r_{k, t - s}(s), & \text{for }  t - k' < s < t  \end{array} \right.$$
as input for the naive prediction of future changes in $r_k$.
The right panel of Table~\ref{tab:lob0} indicates that naive prediction using $x_{k, k'}(t - 1)$ achieves an accuracy of 64.8\%, akin to deep learning outcomes.
This demonstrates that forecasting $r_k$ using deep learning models does not substantially differ from naive predictions.

\subsection{Volatility and directional predictability with a target of $r_{1,20}$}

We now consider a slightly different modified return process that does not contain future information. 
The target variable is
$$ r_{1,k}(t) = \frac{f_k(t) - p_1(t)}{p_1(t)}$$
which calculates the modified return based on a current price $p_1(t)$
as in \textcite{ntakaris2018benchmark}.
Using the model in Table~\ref{tab:model_summary1}, accuracy reaches 54.6\% as in Table~\ref{tab:lob3}. 
It is less accurate than using $r_{20}$ but still surpasses a 33.3\% random prediction.

\begin{table}[t]
	\centering
	\caption{Daily average classification metrics with the entire LOB on $r_{1,20}$}
	\label{tab:lob3}
	\begin{tabular}{c|cccc}
		\hline
		&  Precision & Recall & F1-Score & Size\\
		\hline
		UP &  0.534(0.045) & 0.557(0.086) & 0.541(0.056) & 10,528\\
		DOWN &  0.534(0.051) & 0.558(0.081) & 0.541(0.052) & 10,537\\
		STABLE & 0.548(0.091) & 0.466(0.178) & 0.488(0.139) & 10,465 \\
		\hline
		Accuracy &  \multicolumn{4}{c}{0.546(0.037)}  \\
		\hline
	\end{tabular}
\end{table}

We examine the need for the full LOB dataset, 
focusing on testing the crucial Level 1 data alone.
A less complex model in Table~\ref{tab:model_summary2} is constructed using four-dimensional time series data.

\begin{table}[t]
	\centering	
	\caption{Summary of the simpler model architecture for Level 1 data}	\label{tab:model_summary2}
	\begin{tabular}{c|c|c}
		\hline
		\multicolumn{3}{c}{Input layer with $100 \times 4$ input dimension per data} \\
		\hline
		\multicolumn{3}{c}{Conv2D(32, $1\times 2$, (1, 2))} \\
		\multicolumn{3}{c}{Conv2D(32, $4\times 1$, Zero)} \\
		\multicolumn{3}{c}{Conv2D(32, $4\times 1$, Zero)} \\
		\multicolumn{3}{c}{Conv2D(32, $1\times 2$, (1, 2))} \\
		\multicolumn{3}{c}{Conv2D(32, $4\times 1$, Zero)} \\
		\multicolumn{3}{c}{Conv2D(32, $4\times 1$, Zero)} \\
		\hline
		\multicolumn{3}{c}{Inception module} \\
		\hline
		Conv2D(64, $1\times 1$, Zero) & 
		Conv2D(64, $1\times 1$, Zero) & 
		Max pooling($3\times 1$, Zero) \\
		Conv2D(64, $3\times 1$, Zero)  &	
		Conv2D(64, $5\times 1$, Zero)  &
		Conv2D(64, $1\times 1$, Zero) \\
		\hline
		\multicolumn{3}{c}{Concatenate} \\
		\hline
		\multicolumn{3}{c}{Dropout layer with rate 0.2} \\
		\multicolumn{3}{c}{LSTM layer with 64 units} \\
		\multicolumn{3}{c}{Dense layer with 3 units and softmax function} \\
		\hline
	\end{tabular}
\end{table}

The model with Level 1 data nearly matches the prior model with 53.6\% accuracy.
This suggests that LOB data beyond Level 1 has little impact on mid-price prediction.

\begin{table}[t]
	\centering
	\caption{Daily average classification metrics with Level 1 data on $r_{1,20}$}
	\label{tab:level1}
	\begin{tabular}{c|ccc}
		\hline
		&  Precision & Recall & F1-Score  \\
		\hline
		UP &  0.530(0.042) & 0.542(0.117) & 0.528(0.080) \\
		DOWN & 0.527(0.050) & 0.551(0.096) & 0.533(0.062) \\
		STABLE & 0.526(0.102) & 0.448(0.193) & 0.464(0.146) \\
		\hline
		Overall accuracy & \multicolumn{3}{c}{0.536(0.038)}  \\
		Directional accuracy & \multicolumn{3}{c}{0.711(0.041)} \\
		Volatility accuracy & \multicolumn{3}{c}{0.694(0.081)} \\
		\hline
	\end{tabular}
\end{table}

We now conduct more detailed analysis.
The predictive performance of the above three-class classification problem can be examined in two aspects: 
volatility-based and directional predictability.
Foreseeing if a future return will be STABLE hinges on volatility prediction.
Forecasting an UP or DOWN move when prices deviate from STABLE entails directional prediction.

Let DIVERGE denote the combined UP and DOWN classes. 
Volatility accuracy is the correct STABLE/DIVERGE prediction rate,
while directional accuracy is the UP/DOWN hit ratio within correctly identified DIVERGE cases.
In summary,
\begin{align*}
\text{Volatility accuracy } &=  \frac{\#\text{ of samples that correctly predicted STABLE or DIVERGE}}{\# \text{ of samples}  } \\
\text{Directional accuracy } &= \frac{\# \text{ of samples that correctly predicted UP or DOWN}}{\# \text{ of samples that correctly predicted DIVERGE} }.
\end{align*}

Conceptually, we divide the prediction process into two steps.
First, volatility prediction is conducted to distinguish between STABLE and DIVERGE.
The average accuracy of the volatility prediction is 69.4\% as in the first column in Table~\ref{tab:acc}.
Volatility prediction is well-established in literature \parencite{harvey1992market} and neural network approaches are also extensively studied \parencite{sahiner2023volatility}.

\begin{figure}[t]
	\centering
	\includegraphics[width=0.7\textwidth]{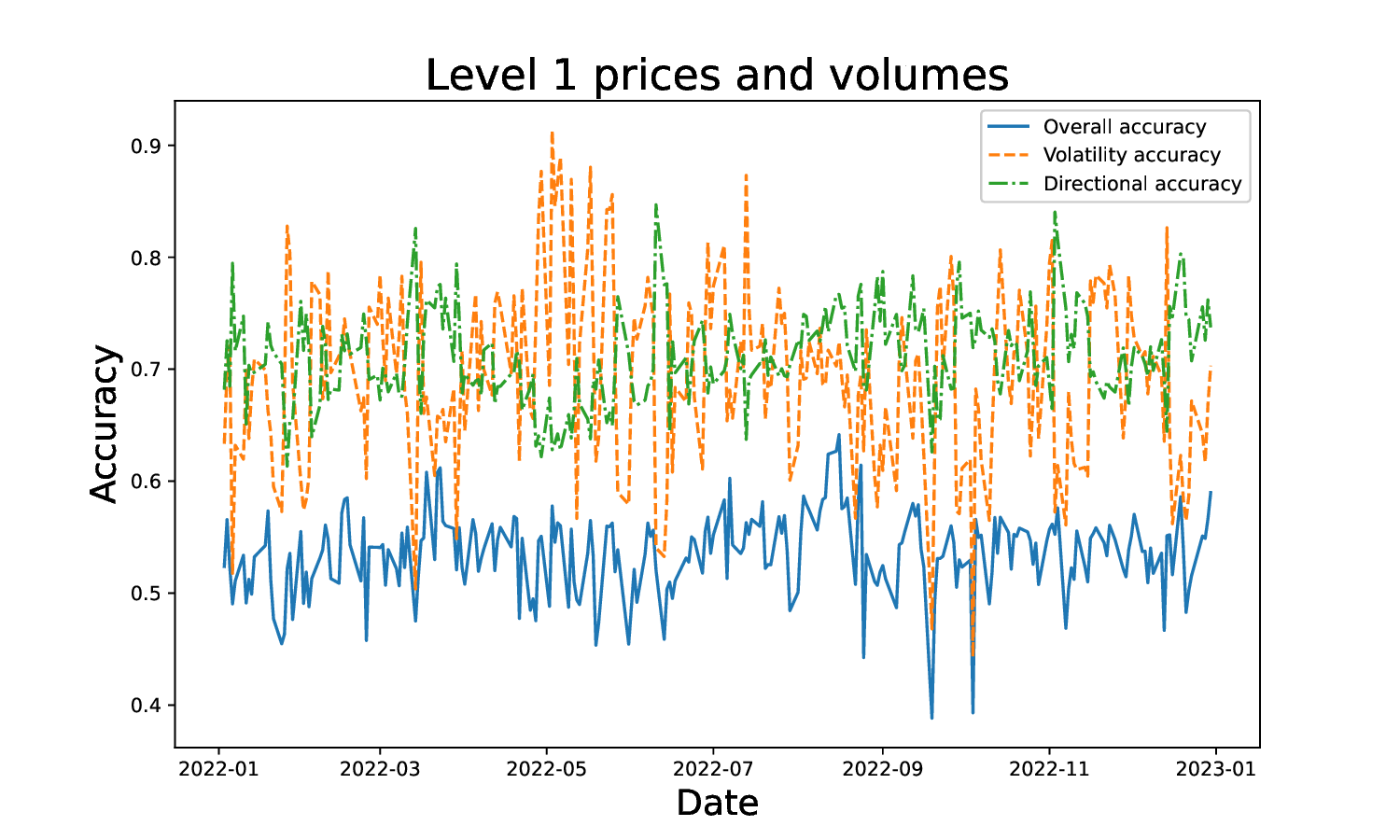}
	\caption{Accuracy metrics based on Level 1 prices and volumes}\label{fig:level1}
\end{figure}

For DIVERGE, the procedure subsequently classifies UP or DOWN,
achieving an average directional prediction accuracy of 71.1\%.
When combined, the overall accuracy is 53.6\%. 
Both volatility and directional predictions significantly exceed the random guess accuracy of 50\%.
Figure~\ref{fig:level1} shows 2022's overall, volatility, and directional accuracy with consistent levels.

\subsection{Level 1 prices or volumes details}

\begin{figure}[t]
	\centering
	\includegraphics[width=0.7\textwidth]{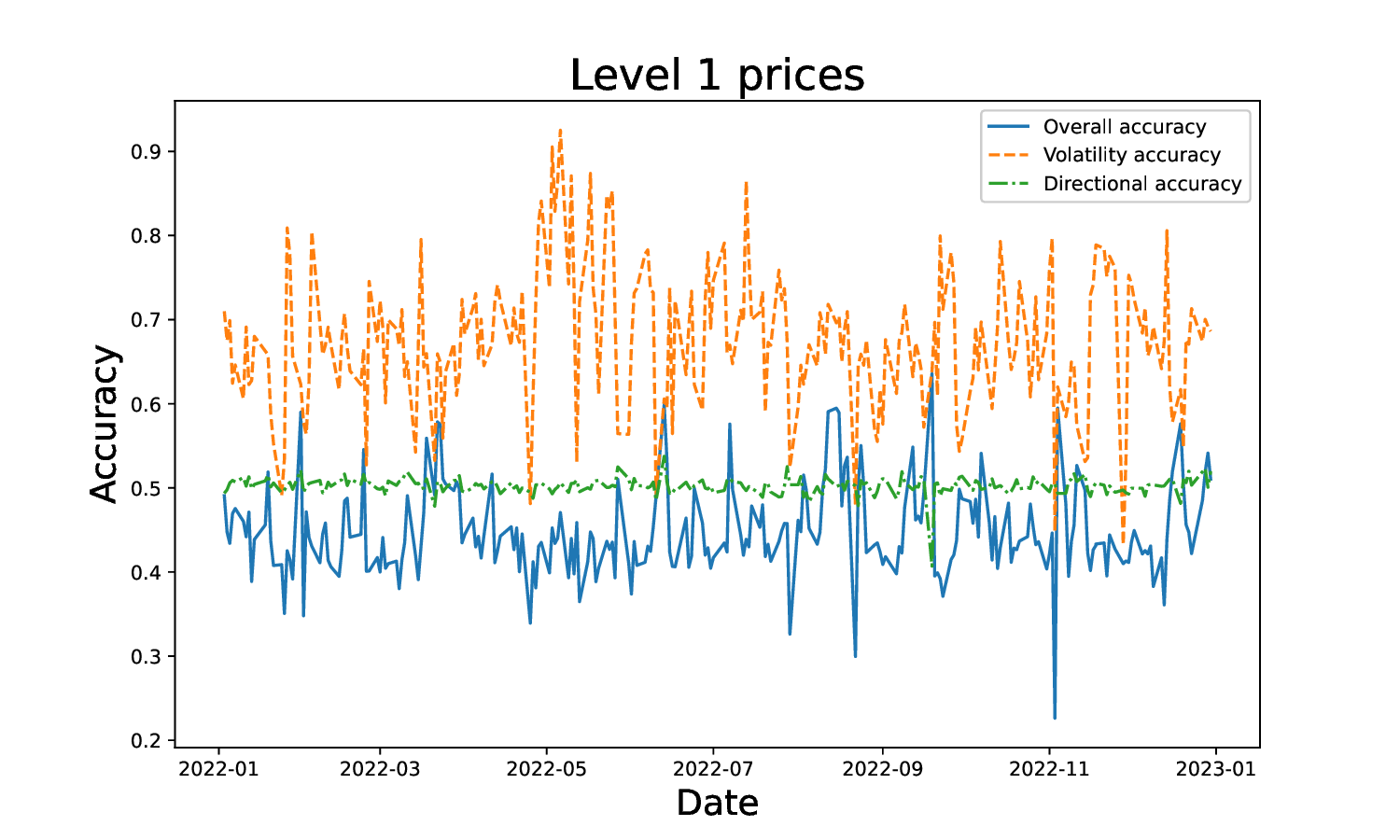}
	\caption{Based on level 1 prices only where 50\% directional accuracy supporting the efficient market hypothesis}\label{fig:level1_price}
\end{figure}

We also examine the case with only Level 1 prices via the simpler model outlined in Table~\ref{tab:model_summary3}.
An interesting observation is that predictions based solely on price processes result in a consistent 50\% directional accuracy, see Figure~\ref{fig:level1_price}.
Relying on price processes alone cannot forecast future price directions
underscoring the efficient market hypothesis~\parencite{fama1970efficient}.

\begin{table}[t]
	\centering	
	\caption{Summary of the model architecture for Level 1 prices or volumes only}	\label{tab:model_summary3}
	\begin{tabular}{c|c|c}
		\hline
		\multicolumn{3}{c}{Input layer with $100 \times 4$ input dimension per data} \\
		\hline
		\multicolumn{3}{c}{Conv2D(32, $1\times 2$)} \\
		\multicolumn{3}{c}{Conv2D(32, $4\times 1$, Zero)} \\
		\multicolumn{3}{c}{Conv2D(32, $4\times 1$, Zero)} \\
		\hline
		\multicolumn{3}{c}{Inception module} \\
		\hline
		Conv2D(64, $1\times 1$, Zero) & 
		Conv2D(64, $1\times 1$, Zero) & 
		Max pooling($3\times 1$, Zero) \\
		Conv2D(64, $3\times 1$, Zero)  &	
		Conv2D(64, $5\times 1$, Zero)  &
		Conv2D(64, $1\times 1$, Zero) \\
		\hline
		\multicolumn{3}{c}{Concatenate} \\
		\hline
		\multicolumn{3}{c}{Dropout layer with rate 0.2} \\
		\multicolumn{3}{c}{LSTM layer with 64 units} \\
		\multicolumn{3}{c}{Dense layer with 3 units and softmax function} \\
		\hline
	\end{tabular}
\end{table}

Meanwhile, volatility-based predictions exhibit high 67.5\% accuracy.
This aligns with studies that indicate future volatility can be predicted based on past prices.
In summary, Level 1 volumes are vital for directional forecasts and marginally enhance volatility prediction.
We also train and predict using only Level 1 volumes, as shown in the fourth column of Table~\ref{tab:acc}.

\begin{table}[t]
	\centering
	\caption{Accuracy comparison across different data sources for Level 1 data}
	\label{tab:acc}
	\begin{tabular}{c|c|cccc}
		\hline
		& Accuracy &  Prices \& volumes & Prices & Volumes & Prices \& imbalance \\
		\hline
		     & Overall &  0.536(0.038) & 0.448(0.053) & 0.464(0.085) & 0.517(0.036)\\
		AAPL & Directional &  0.711(0.041)& 0.503(0.010) & 0.589(0.154) & 0.705(0.041)\\
		     & Volatility &  0.694(0.081) & 0.675(0.082) & 0.681(0.080) & 0.681(0.082)\\
		\hline
		     & Overall & 0.504(0.077) & 0.459(0.099) & 0.464(0.114) & 0.493(0.070) \\
		AMZN & Directional & 0.615(0.060) & 0.502(0.018) & 0.534(0.099) & 0.611(0.067)\\
		     & Volatility &  0.674(0.117) & 0.682(0.122) & 0.626(0.118) & 0.665(0.116)\\
		\hline
		     & Overall & 0.468(0.114) & 0.415(0.127) & 0.394(0.128) & 0.461(0.112)\\
		MSFT & Directional & 0.636(0.061) & 0.499(0.066) & 0.550(0.111) & 0.632(0.056)\\
		     & Volatility & 0.676(0.170) & 0.645(0.193) & 0.618(0.176) & 0.667(0.175)  \\
		\hline
		     & Overall &  0.465(0.070) & 0.421(0.091) & 0.392(0.074) & 0.456(0.072)\\
		NVDA & Directional & 0.619(0.066) & 0.501(0.041) & 0.528(0.082) & 0.615(0.067)\\
		     & Volatility & 0.689(0.150) & 0.670(0.162) & 0.691(0.170) & 0.681(0.155)\\
		\hline
	\end{tabular}
\end{table}

The results emphasize that both price and volume are key for optimal prediction.
Price data alone is inadequate for directional forecasting.
This is related to research by \textcite{kamalov2021financial}, \textcite{gurrib2022predicting} and \textcite{ma2023reliability}, which shows that incorporating other factors like news and technical indicators can enhance directional price prediction.

To investigate the factors contributing to improved predictions, 
we use volume imbalance as an input,
which have long been considered a factor affecting future returns
\parencite{chordia2004order}, in addition to the add bid/ask prices.
The result almost matches the performance achieved when utilizing all Level 1 data as shown in Table~\ref{tab:acc}.
The volume information affecting the prediction is mostly the volume imbalance.
The same analysis on AMZN, NVDA, and MSFT in 2022 yields similar results.

Can we obtain low risk profits by predicting price direction through volume imbalance? 
When the volume is uneven,  price is likely to move towards the weaker side. 
However, profiting through directional speculation faces challenges, 
as obtaining the desired orders on the strong side amid tough competition is required. 
Thus, a strategy aimed at low risk profits through directional prediction may have limitations,
again suggesting the validity of the efficient market hypothesis.

\section{Conclusion}

We explored the predictive power of the mid-price process using AAPL's 2022 LOB data and deep learning models based on the modified return process.
Using only Level 1 data produced results comparable to those of the entire LOB dataset.
Predicting future price changes into UP, DOWN, and STABLE categories involves directional and volatility predictions.
Using Level 1 data, both predictions achieved an accuracy of approximately 70\%. 
However, using only price information resulted in 50\% accuracy for directional prediction.
Incorporating volume information, especially volume imbalance, significantly enhanced directional prediction.

\section*{Acknowledgements}
This work was supported by the 2024 Yeungnam University Research Grant.

\section*{Disclosure of Interest}
No potential conflict of interest was reported by the author(s).


\printbibliography 

\end{document}